# On the question of radiative losses in the frame of DM formalism.

## Part I: Free space

### A. Chipouline

1. Introduction

2. Dynamics of classic and quantum dipole
    2.1 Dynamics of classic dipole
    2.2 Dynamics of quantum dipole

3. Math formalism for radiative losses and relaxation dynamics
    3.1 Model formulation
    3.2 Stationary dynamics
    3.3 Relaxation dynamics

4. Radiative losses for classic and quantum dipoles in free space
    4.1 Stationary state in free space
    4.2 Relaxation dynamics in free space

5. Conclusions

6. Appendixes
    6.1 Appendix 1: density matrix vs harmonic oscillator equations
    6.2 Appendix 2: Maxwell equations & density matrix formalism

*************************************************************************

**Abstract**

It is shown that the description of the radiative losses in the case of density matrix (DM) equations requires special attention. A widely accepted harmonic oscillator equation cannot be used to describe dynamics of a quantum emitter: both approaches (harmonic oscillator equations and DM) give different results. Moreover, a formal substitution of a harmonic oscillator equation (classic dipole) by the DM equations leads to unphysical results. This paradox appears in the case of description of e.g. relaxation dynamics (including Purcell effect) and spaser dynamics in the case of multimode operation. This paradox can be resolved by careful subdivision of the considered fields by regular and stochastic ones, which is actually a prerequisite for the DM equations. Peculiarities of the extraction of the relaxation constants are discussed.

# 1. <u>Introduction</u>

The problem of radiative losses of classic or quantum emitters is one of the central at the consideration of spaser/nanolaser (more generally coupled plasmonic nano-particle NP and quantum system QS – NP/QS) stationary as well as relaxation dynamics. Nevertheless, this problem has not received special attention and up to now often is a source of ambiguous interpretations.

The classical consideration of the radiative losses for a free dipole (with dynamics described by in the frame of harmonic oscillator equation) leads to a well-known extra term in the dynamic equation proportional to the third time derivative [XXX]. Nevertheless, this simple dynamic equation modification is justified for the case of free space only. In case of the coupled dynamics (e.g. spaser/nanolaser) the radiative losses do not possess this straightforward modelling.

The classic harmonic oscillator model has been widely used in the vast majority of publications devoted to the coupled dynamics of an emitter and plasmonic nanoresonator [Novotny 06]. This approximation (classic harmonic oscillator instead of quantum DM formalism) actually cannot be used in case of significant inversion variations; this question is considered in Appendix 1. In spite of this trivial textbook fact, the harmonic oscillator model for quantum emitter dynamics is still surprisingly accepted in articles, see recent publications [Poddubny 13], [Decker 13], [Sauvan 13], and review [Agio 13].

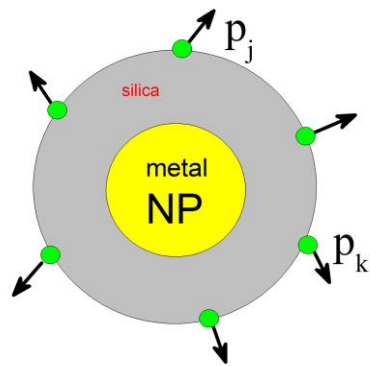

**Fig. 1: Scheme of the considered object**

The simply use of DMformalism instead of harmonic oscillator one (e.g. to describe spaser dynamics [Stockman 10]) lead to a paradox considered in this paper. The essence of this paradox is rooted in the fact, that the relaxation processes are in fact considered in this case doubly: first by phenomenological energy and phase relaxation times, and second at the consideration of the radiative losses. Appearance of this paradox shows that the problem of dynamics of a quantum emitter coupled with a plasmonic nanoresonator requires fundamentally more accurate treatment.

We introduce a particular system, which will be kept in mind hereafter. The system consists of nanoresonator, which will be called nanoparticle NP, coupled with quantum emitters (Quantum System – QS) as it is shown in Fig. 1.

**2. Dynamics of classical and quantum dipole**

**2.1 Dynamics of classical dipole**

The dynamics of a QS in the frame of the harmonic oscillator (HO) equation is as usual:

$$\frac{d^2 p}{dt^2} + 2\gamma \frac{dp}{dt} + \omega_0^2 p = \chi A \exp(-i\omega t) + F \exp(-i\omega t) \tag{2.1}$$

The external force $F$ brings energy into the system, it can be an external filed (incoming wave). This external force $F$ is to some extend equivalent to the pump in the case of quantum dipole, described by a density matrix.

Here, an energy relaxation channel, described by $\gamma$ is presented. This channel takes into account non radiative relaxation of energy from HO to a thermo bath. In spite of the fact that equation (2.1) is probably mostly used equation in physics, the nature of $\gamma$ and rigorous elaboration of this term is not widely discussed. It is worth to recall it in order to avoid misinterpretation in future discussion. The model described by (2.1) consists of a single oscillator weakly coupled to a lot of similar oscillators, which are in turn weakly coupled with each other. This system of huge number of weakly coupled HOs is called thermo bath or thermostat. The thermo bath is one of the widely used models appearing from statistical quantum physics. The qualitative picture behind this model is also known from the university courses of physics: excitation in HO gets transferred to other coupled oscillators, which in turn give energy further to other ones and thus far the initially localized energy (oscillation of a single dipole) gets diffused, or, in other words, thermalized among the other oscillators. Inverse energy transfer is prohibited by the first law of thermodynamics: the dissipated (diffused) energy cannot come back to an origin (i.e. single oscillator under consideration). It is also important to realize, that all the oscillators (or, more generally, eigen modes) at the nonzero temperature exhibit stochastic oscillations with arbitrary phase. Hamiltonian of a single HO interacting with thermostat consists of "eigen" part $H_0$ giving eigen states, regular interaction $V_R$, and stochastic interaction $V_{int}$:

$$\begin{cases} H = H_0 + V_R + V_{stoch}, \\ \langle V_R \rangle = V_R, \\ \langle V_{stoch} \rangle = 0 \end{cases} \quad (2.2)$$

"Eigen" part $H_0$ leads to dynamic part $\frac{d^2 p}{dt^2} + \omega_0^2 p$, regular interaction $V_R$ is responsible for the right side of (2.1) $\chi A \exp(-i\omega t) + F \exp(-i\omega t)$, while stochastic part $V_{int}$ results (using iterative approach) in relaxation term $2\gamma \frac{dp}{dt}$ in the second order of approximation of the stochastic interaction: the first order gives evidently zero results. The radiative losses in this case is the consequence of regular interaction, namely $\chi A \exp(-i\omega t)$ in (2.1) and cannot be included in $\gamma$.

In Fourie space (2.1) becomes:

$$\begin{cases} \left(\omega_0^2 - \omega^2 - 2i\gamma\omega\right) p = \chi A + F \\ p = \frac{\chi A + F}{L_\omega} = \frac{\chi A + F}{\omega_0^2 - \omega^2 - 2i\gamma\omega} = \beta_c(\omega)(\chi A + F) \\ \beta_c(\omega) = \frac{1}{\omega_0^2 - \omega^2 - 2i\gamma\omega} \\ \text{Re}\left[\beta_c(\omega)\right] = \frac{\omega_0^2 - \omega^2}{\left(\omega_0^2 - \omega^2\right)^2 + 4\gamma^2\omega^2} \\ \text{Im}\left[\beta_c(\omega)\right] = \frac{2\gamma\omega}{\left(\omega_0^2 - \omega^2\right)^2 + 4\gamma^2\omega^2} \\ \left|\beta_c(\omega)\right|^2 = \frac{1}{\left(\omega_0^2 - \omega^2\right)^2 + 4\gamma^2\omega^2} \end{cases} \quad (2.3)$$

which describes stationary amplitudes *p* and *A*. Here *A* in the first equation (2.3) is the "self-action" field generated by a dipole at its origin.

## 2.2 Dynamics of quantum dipole

Quantum dynamics of the radiating point dipole is considered now in the frame of the DM approach. The dynamics is given by well-known system of equations with artificial pump in SVA (Slowly Varying Approximation) representation [Chipouline 12 JOPT]:

$$\begin{cases} \dfrac{dp}{dt} + p\left(\dfrac{1}{\tau_2} + i(\omega - \omega_{21})\right) = \dfrac{i\mu^2 A^* n}{\hbar} \\ \dfrac{dn}{dt} + \dfrac{(n - n_0)}{\tau_1} = \dfrac{i(Ap - A^* p^*)}{2\hbar} \\ n = \rho_{22} - \rho_{11} \\ p = \mu \rho_{12} \\ n_0 = \dfrac{(W\tilde{\tau}_1 - 1)}{(W\tilde{\tau}_1 + 1)} \end{cases} \quad (2.4)$$

Here $\rho_{22}$, $\rho_{11}$ and $\rho_{12}$, $\rho^*_{12}$ are the diagonal and non-diagonal matrix density elements, respectively; $\tau_2$ and $\tilde{\tau}_1$ are the constants describing phase and energy relaxation processes due to the interaction with a thermostat; $\omega_{21} = (E_2 - E_1)/\hbar$ is the transition frequency between levels 2 and 1; $H_{12}$ is the Hamiltonian matrix element responsible for interaction with the external fields; *W* is the phenomenological pump rate.

First of all at the elaboration of the DM approach (and consequently (2.4)) the same subdivision by regular and stochastic interaction (2.2) and all respective math for the elaboration of (2.4) has been accepted [Fano 57], [Fain 72], [Akulin 87]. Equations (2.4) are written for already averaged values and do not contain stochastic parts − all stochastic processes (interaction with thermo bath and zero field fluctuations) are packed in the two relaxation times $\tau_1$ and $\tau_2$, exactly as the stochastic interaction with the thermo bath has been packed into $\gamma$ in (2.1). The regular part of the Hamiltonian of interaction, again in full analogy with (2.1) results in right side of (2.4), namely $\dfrac{i\mu^2 A^* n}{\hbar}$ and $\dfrac{i(Ap - A^* p^*)}{2\hbar}$.

## 3. Math formalism for coupled dynamics with radiative losses

### 3.1 Model formulation

The problem of the dynamics of an emitter in free space or near some nanoobject is one of the fundamental and has been considered in textbooks [Novotny 06]. Mathematically, the problem is formulated by a system of coupled equations between dynamics of dipole (in quantum or classical approximations) and Helmholtz equation for the electromagnetic field. We assume here for classical dipole dynamics an accepted in most of publications HO equation and DM equations for two level system with an artificial pump. The final systems are:

$$\begin{cases} \dfrac{d^2 p_i}{dt^2} + 2\gamma \dfrac{dp_i}{dt} + \omega_0^2 p_i = \chi A(r_i,t)\exp(-i\omega t) + F(r_i,t)\exp(-i\omega t) \\ \Delta E(r,t) - \dfrac{1}{c^2}\dfrac{\partial^2 D(r,t)}{\partial t^2} = -\dfrac{4\pi}{c^2}\dfrac{\partial^2}{\partial t^2}\sum_i p_i(r_i,t)\delta(r-r_i) \end{cases} \quad (3.1)$$

$$\begin{cases} \dfrac{dp_i}{dt} + p_i\left(\dfrac{1}{\tau_{2,i}} + i(\omega - \omega_{21})\right) = \dfrac{i\mu^2 A(r_i,t)^* n}{\hbar} \\ \dfrac{dn_i}{dt} + \dfrac{(n_i - n_0)}{\tau_{1,i}} = \dfrac{i\left(A(r_i,t)p_i - A(r_i,t)^* p_i^*\right)}{2\hbar} \\ n_i = \rho_{22,i} - \rho_{11,i} \\ p_i = \mu_i \rho_{12,i} \\ n_{0,i} = \dfrac{(W_i \tilde{\tau}_1 - 1)}{(W_i \tilde{\tau}_1 + 1)} \\ \Delta E(r,t) - \dfrac{1}{c^2}\dfrac{\partial^2 D(r,t)}{\partial t^2} = -\dfrac{4\pi}{c^2}\dfrac{\partial^2}{\partial t^2}\sum_i p_i(r_i,t)\delta(r-r_i) \end{cases} \quad (3.2)$$

Here function $F$ in classic version (3.1) is the external force (can be an external electric field as well), $p_i$ – are the dipole moments of the active molecules, $D$ – is the dielectric displacement function, taking into account a possible presence of the nanoobjects. Particular physical picture behind (3.1) is as usual: coupled equations for the field and polarizability of the emitters $p_i$. It is assumed that $p_i$ are the dipole moments (in the case of zero dipole moment, $p_i$ is the qudrupole and etc. higher moments, which is rear). In this paper dipole moment is assumed to be nonzero and dominates as a source of radiation for $p_i$. At the same time, a nanoresonator in the vicinity of the emitter can be taken into account by $D$ with all possible eigen modes [Pustovit 10].

It is important to note that the field *E* in Helmholtz equation includes all possible fields: incoming (if any), scattered, and emitted by $p_i$. The radiative losses are part of this field and are taken into account in (3.1), but NOT included in $\gamma$.

As for the radiative losses in (3.2), the consideration in this case is much less trivial. Remind that in the case of the HO equation (3.1) the radiative losses are not included due to the fact that the radiative losses are caused not by stochastic part of the interaction Hamiltonian, but rather by the regular one – see (2.7), where there is only one stochastic coefficient $\gamma$. In the case of quantum consideration, there is a purely quantum contribution to the stochastic part of the interaction, namely interaction with vacuum field fluctuations. They are packed, in addition to the non-radiative relaxation to thermo bath, into the relaxation times $\tau_1$ and $\tau_2$, e.g. for the energy relaxation time $\tau_1$:

$$\frac{1}{\tau_1} = \frac{1}{\tau_{1,r}} + \frac{1}{\tau_{1,nr}} \qquad (3.3)$$

Here $\tau_{1,r}$ and $\tau_{1,nr}$ are the radiative and non-radiative relaxation times respectively. In fact, the radiative relaxation time is given by the well-known expression [Novotny 06]:

$$\frac{1}{\tau_{1,r}} = \frac{2\omega}{3\hbar\varepsilon_0}|\mu|^2 \rho_\mu\left(\vec{r_0},\omega_0\right)$$

$$\rho_\mu\left(\vec{r_0},\omega_0\right) = 3\sum_k \left[\vec{n_\mu} \cdot \left(\vec{u_k}\vec{u_k^*}\right) \cdot \vec{n_\mu}\right]\delta(\omega_k - \omega_0) \qquad (3.4)$$

Here $\rho_\mu\left(\vec{r_0},\omega_0\right)$ is the local density of states (LDOS), which depends on the presence/absence of any objects in the vicinity of the quantum emitter, e.g. LDOS will be modified in the case of coupling between NP and $p_i$, see Fig. 1.

Now the question is: do we need to consider radiative losses the same way like for HO (3.1), or there are some differences? In order to answer this question, we elaborate solutions in parallel for the HO (3.1) and DM (3.2) models.

We are interested in two operation modes, namely stationary states and relaxation dynamics.

### 3.2 Stationary state

The stationary dynamics assumes that the system parameters stationary oscillate at some frequency ω, where the frequency is supposed to be real value; an exact numerical value for this frequency can be found from (3.1), (3.2) depends on particular conditions. We introduce slowly varying approximation as usual:

$$\begin{cases} p_i(r_i,t) = p_i(r_i,\omega)\exp(-i\omega t) \\ E(r,t) = A(r,\omega)\exp(-i\omega t) \\ D(r,t) = D(r,\omega)\exp(-i\omega t) \\ \Delta A(r,\omega) + \dfrac{\omega^2}{c^2}D(r,\omega) = \dfrac{4\pi\omega^2}{c^2}\sum_i p_i(r_i,\omega)\delta(r-r_i) \end{cases} \quad (3.5)$$

It has to be pointed out that in principle we can assume Laplace transformation instead of Fourie, but in this case the next step, expression $D$ through the dielectric constant and $E$, needs to be carefully considered. The dielectric function $\varepsilon(\omega)$ is the Fourie image of the time dependent dielectric function with all properties implied by causality/passivity principles; Laplace image of the dielectric function is not usually considered.

Assuming:

$$D(r,\omega) = \varepsilon(r,\omega)A(r,\omega) \quad (3.6)$$

we get:

$$\begin{cases} p_i = \dfrac{\chi A(r_i,\omega) + F(r_i,\omega)}{L_\omega} = \dfrac{\chi A(r_i,\omega) + F(r_i,\omega)}{\omega_0^2 - \omega^2 - 2i\gamma\omega} = \beta_c(\omega)\big(\chi A(r_i,\omega) + F(r_i,\omega)\big) \\ \beta_c(\omega) = \dfrac{1}{\omega_0^2 - \omega^2 - 2i\gamma\omega} \\ \Delta A(r,\omega) + \dfrac{\omega^2}{c^2}\varepsilon(r,\omega)A(r,\omega) = \dfrac{4\pi\omega^2}{c^2}\sum_i p_i(r_i,\omega)\delta(r-r_i) \end{cases} \quad (3.7)$$

Only single QS is assumed in order to demonstrate the paradox, extension on multiple QS is straightforward.

The Helmholtz equation allows us to consider any shapes and layouts of NPs, consideration of a single NP and single molecule is stipulated by the fact, that the problem in this case possesses an analytical treatment and at the same time pronouncedly shows the physics behind.

The main step in the transformation of the Helmholtz equation is in the design of the appropriate Green function, taking into account the presence of the NP. It means effectively, that we replace our space with the NP with another space without this NP, but with a more sophisticated Green function. After that, the last equation can be trivially rewritten as:

$$\Delta A(r,\omega) + \frac{\omega^2}{c^2}\varepsilon(r,\omega)A(r,\omega) = \frac{4\pi\omega^2}{c^2}p(r_p,\omega)\delta(r-r_p)$$
$$\Downarrow$$
$$A(r,\omega) = \frac{4\pi\omega^2}{c^2}G(r,r_p,\omega)p(r_p,\omega)$$
(3.8)

Finally, the problem for stationary states is formulated by the master system of equations for classical case:

$$\begin{cases} p = \beta_c(\omega)\left(\chi A_{NF}(r_p,\omega) + F(r_p,\omega)\right) \\ A_{NF}(r,\omega) = \frac{4\pi\omega^2}{c^2}G(r,r_p,\omega)p(r_p,\omega) \end{cases}$$
(3.9)

and for the quantum case:

$$\begin{cases} p\left(\frac{1}{\tau_2} + i(\omega-\omega_{21})\right) = \frac{i\mu^2 A_{NF}^*(r_p,\omega)n}{\hbar} \\ \frac{(n-n_0)}{\tau_1} = \frac{i\left(A_{NF}(r_p,\omega)p - A_{NF}^*(r_p,\omega)p^*\right)}{2\hbar} \\ A(r,\omega) = \frac{4\pi\omega^2}{c^2}G(r,r_p,\omega)p(r_p,\omega) \end{cases}$$
(3.10)

It is worth noting again, that (3.9) and (3.10) describe the possible stationary states and does not give any information about transient system dynamics.

In addition to the losses in QS itself described by $\gamma$ (again, radiative losses are not included in $\gamma$), there are irreversible losses described by an imaginary part of the dielectric constant. The physical picture of these losses (ohmic losses) is in the interaction of free electron with the solid state lattice resulted in transfer of the kinetic energy of electron into the elementary collective lattice excitations e.g. phonons. Radiative losses are NOT included in the imaginary part of epsilon.

### 3.3 Relaxation dynamics

Investigation of the relaxation dynamics assumes that we are looking for the solution of (3.1) and (3.2) in time domain. Typical relaxation process assumes that we do not have during the process an extra energy delivery into the system: the system starts at the zero time from some nonzero values and relaxes to zero levels of the respective variables. In the case of the linear system, it would be reasonable to assume that the relaxation follows some exponential function. It means that we can use the same substitution (2.3) but the frequency $\omega$ in this case becomes complex value. The imaginary part of this value gives the relaxation time, while real part describes frequency shift; both are supposed to be found from master systems of equations (3.9) and (3.10) with an extra assumption of complex nature of $\omega$.

## 4. Radiative losses for classic and quantum dipole in free space
### 4.1 Stationary state in free space

Results for the $p$ and $A_{NF}$ from (3.9) are:

$$\begin{cases} p = \dfrac{\beta_c(\omega) F}{\left[1 - \dfrac{4\pi\omega^2 \chi}{c^2} \beta_c(\omega) G_{NF}\right]} = \beta_{c,eff}(\omega) F \\ A_{NF} = \dfrac{4\pi\omega^2}{c^2} \dfrac{\beta_c(\omega) G_{NF} F}{\left[1 - \dfrac{4\pi\omega^2 \chi}{c^2} \beta_c(\omega) G_{NF}\right]} = \dfrac{4\pi\omega^2}{c^2} \beta_{c,eff}(\omega) G_{NF} F \\ \beta_{c,eff}(\omega) = \dfrac{\beta_c(\omega)}{\left[1 - \dfrac{4\pi\omega^2 \chi}{c^2} \beta_c(\omega) G_{NF}\right]} \end{cases} \quad (4.1)$$

Amplitude, intensity, and total radiative power in the far field zone can be straightforwardly calculated using the known Green function for far field:

$$\begin{cases} A_{FF} = \dfrac{4\pi\omega^2}{c^2} \dfrac{\beta_c(\omega) G_{FF} F}{\left[1 - \dfrac{4\pi\omega^2 \chi}{c^2} \beta_c(\omega) G_{NF}\right]} = \dfrac{4\pi\omega^2}{c^2} \beta_{c,eff}(\omega) G_{FF} F \\ I_{FF} = \dfrac{c}{8\pi} \left(\dfrac{4\pi\omega^2}{c^2}\right)^2 |G_{FF} p|^2 = \dfrac{2\pi\omega^4}{c^3} \dfrac{|\beta_c(\omega) G_{FF}|^2}{\left|1 - \left(\dfrac{4\pi\omega^2 \chi}{c^2}\right) \beta_c(\omega) G_{NF}\right|^2} |F|^2 = \dfrac{2\pi\omega^4}{c^3} |\beta_{c,eff}(\omega) G_{FF}|^2 |F|^2 \\ P_{FF} = \dfrac{c}{8\pi} \left(\dfrac{4\pi\omega^2}{c^2}\right)^2 \oint |G_{FF} p|^2 dS = \dfrac{8\pi^2 \omega^4}{c^3} |\beta_{c,eff}(\omega)|^2 |F|^2 = \dfrac{8\pi^2 \omega^4}{c^3} \dfrac{|\beta_c(\omega)|^2 |F|^2}{\left|1 - \dfrac{4\pi\omega^2 \chi}{c^2} \beta_c(\omega) G_{NF}\right|^2} \end{cases}$$
(4.2)

Systems (4.1), (4.2) are universal. It means that the systems describe stationary state for any possible Green functions, i.e. for any possible nanoobject which can be described by $\varepsilon(r,\omega)$. Let us consider first the problem of the radiative losses in free space. In order to complete (4.2) we have to know Green function for free space in the near field and far field zones, which are [Pustovit 10]:

$$\begin{cases} G_{NF} = \dfrac{c^2}{4\pi\omega^2} \dfrac{2}{3} ik^3 \\ G_{FF} = \dfrac{1}{r} \exp(ikr) \end{cases}$$
(4.3)

here $G_{NF}$ and $G_{FF}$ are the Green function of the free space in the near field and far field zones respectively. The oscillating dipole is driven by the force $F$ from one side (energy income) and loses energy into thermo bath and radiation. If the energy losses are compensated by the external force $F$, the system reaches the stationary state. It is important to understand that here Green function for the near field has to be taken, hence $A_{NF} = \dfrac{4\pi\omega^2}{c^2} G_{NF} p$. In this chapter we are interested in free space (no nanoobjects in near field zone). Let us substitute (4.3) into (4.2). The total emitted power becomes:

$$\begin{cases} p = \dfrac{\beta_c(\omega) F}{\left[1 - \dfrac{2i\chi}{3}\beta_c(\omega)\left(\dfrac{\omega}{c}\right)^3\right]} \\[2ex] P_{FF,C}(\omega) = \dfrac{8\pi^2 \omega^4}{c^3} \dfrac{1}{\left((\omega_0^2 - \omega^2)^2 + 4\gamma^2\omega^2\right)} \dfrac{|F|^2}{\left(1 + \dfrac{2}{3}\left(\dfrac{\omega}{c}\right)^3 \dfrac{\chi\gamma\omega}{(\omega_0^2-\omega^2)^2 + 4\gamma^2\omega^2}\right)^2 + \left(\dfrac{2}{3}\left(\dfrac{\omega}{c}\right)^3 \dfrac{\chi(\omega_0^2-\omega^2)}{(\omega_0^2-\omega^2)^2 + 4\gamma^2\omega^2}\right)^2} \\[2ex] P_{FF,C}(\omega_0) = \dfrac{2\pi^2 \omega_0^2}{\gamma^2 c^3} \dfrac{|F|^2}{\left(1 + \dfrac{\chi\omega_0^2}{6c^3\gamma}\right)^2} \end{cases} \quad (4.4)$$

which clearly indicates that the emitted power becomes lower if the radiative losses (near field Green function) are taken into account. Effective dipole has been introduced in (4.1), which is characterized by its own spectral response function $\beta_{c,eff}(\omega)$. The respective dynamic equation for the dipole with the radiative losses is governed by:

$$\beta_{c,eff}^{-1}(\omega) = \beta_c^{-1}(\omega)\left[1 - \dfrac{2i}{3}\chi\beta_c(\omega)k^3\right] = \beta_c^{-1}(\omega) - \dfrac{2}{3}\chi k^3 = \beta_c^{-1}(\omega) - \dfrac{2i}{3}\chi\left(\dfrac{\omega}{c}\right)^3 \quad (4.5)$$

which corresponds to the extra term proportional to a third derivative in dynamic equation (2.1):

$$\dfrac{d^2 p}{dt^2} + 2\gamma \dfrac{dp}{dt} + \omega_0^2 p + \dfrac{2}{3}\dfrac{\chi}{c^3}\dfrac{d^3 p}{dt^3} = F\exp(-i\omega t) \quad (4.6)$$

**The radiative losses can be equivalently taken into account by an extra term in the dipole dynamic equation. In this case, the dipole oscillation amplitude *p* has to be taken from equation (4.6), not from (2.1).**

In text books, the final expressions for the output power (4.5) are usually integrated over the whole spectrum, which is not done here – the final expressions for the integrated power are not observable and are not suited for the analytical consideration. It nevertheless can be done pretty easy using any available numerical package.

Following the same logic as for classic dipole (3.9), consider now a stationary state of quantum oscillator (3.10). Analog of (4.2) is:

$$\begin{cases} p\left(\dfrac{1}{\tau_2}+i(\omega-\omega_{21})\right)=\dfrac{i\mu^2 A_{NF}^* n}{\hbar} \\ \dfrac{(n-n_0)}{\tau_1}=\dfrac{i\left(A_{NF}p - A_{NF}^* p^*\right)}{2\hbar} \\ A_{NF}=\dfrac{4\pi\omega^2}{c^2}G_{NF}p^* \end{cases} \qquad (4.7)$$

Substituting $A_{NF}$ from the last equation to the other two, we have:

$$\begin{cases} p\left(1+i\tau_2(\omega-\omega_{21})-\dfrac{i\mu^2\tau_2 n}{\hbar}\dfrac{4\pi\omega^2}{c^2}G_{NF}^*\right)=0 \\ n=n_0-\dfrac{\tau_1}{\hbar}\dfrac{4\pi\omega^2}{c^2}\operatorname{Im}[G_{NF}]|p|^2 \end{cases} \qquad (4.8)$$

Along with the trivial solution $p=0$, (4.8) possesses a nontrivial one given by:

$$\begin{cases} 1+i\tau_2(\omega-\omega_{21})=\dfrac{i\mu^2\tau_2 n}{\hbar}\dfrac{4\pi\omega^2}{c^2}G_{NF}^* \\ n=n_0-\dfrac{\tau_1}{\hbar}\dfrac{4\pi\omega^2}{c^2}\operatorname{Im}[G_{NF}]|p|^2 \end{cases} \qquad (4.9)$$

which in turn is reduced to:

$$\begin{cases} \dfrac{\hbar(1+i\tau_2(\omega-\omega_{21}))}{i\mu^2\tau_2\dfrac{4\pi\omega^2}{c^2}G_{NF}^*}=n \\ n=n_0-\dfrac{\tau_1}{\hbar}\dfrac{4\pi\omega^2}{c^2}\operatorname{Im}[G_{NF}]|p|^2 \end{cases} \qquad (4.10)$$

The first equation gives stationary frequency $\omega_{eigen}$ from the condition $Im[n]=0$, and then consequently $n(\omega_{eigen})$. From the second equation we get then stationary value of modulus $p$ square:

$$\begin{cases} Im\left[\dfrac{\hbar(1+i\tau_2(\omega-\omega_{21}))}{i\mu^2\tau_2\dfrac{4\pi\omega^2}{c^2}G_{NF}^*}\right]=0 \Rightarrow \omega_{eigen} \\ Re\left[\dfrac{\hbar(1+i\tau_2(\omega-\omega_{21}))}{i\mu^2\tau_2\dfrac{4\pi\omega^2}{c^2}G_{NF}^*}\right]=n \\ |p|^2=\dfrac{\hbar}{\tau_1\dfrac{4\pi\omega^2}{c^2}Im[G_{NF}]}\left(n_0-Re\left[\dfrac{\hbar(1+i\tau_2(\omega-\omega_{21}))}{i\mu^2\tau_2\dfrac{4\pi\omega^2}{c^2}G_{NF}^*}\right]\right) \end{cases} \quad (4.11)$$

which is supposed to be an analog of (4.1) for classical dipole. Remember, that the eigen values did not appear at all at the consideration of the classical HO. It reflects the principal difference: a HO model does not assume pumping, but rather driving by an external force. It brings us to the conclusion, that the quantum case is basically closer to a nonlinear oscillator: in both cases the dipole dynamics is nonlinear and possesses an eigen stationary state (will be considered elsewhere).

Stationary state for free space is given by a substitution of the respective near field Green function (4.3) into (4.11):

$$\begin{cases} Im\left[\dfrac{\hbar(1+i\tau_2(\omega-\omega_{21}))}{\mu^2\tau_2\dfrac{2}{3}\left(\dfrac{\omega}{c}\right)^3}\right]=0 \Rightarrow \omega_{eigen}=\omega_{21} \\ Re\left[\dfrac{\hbar(1+i\tau_2(\omega-\omega_{21}))}{\mu^2\tau_2\dfrac{2}{3}\left(\dfrac{\omega}{c}\right)^3}\right]=n \Rightarrow n=\dfrac{3\hbar c^3}{2\mu^2\tau_2\omega^3} \\ |p|^2=\dfrac{2\hbar c^3}{3\tau_1\omega^3}\left(n_0-\dfrac{3\hbar c^3}{2\mu^2\tau_2\omega^3}\right) \end{cases} \quad (4.12)$$

With the following numerical values:

$$\begin{cases} \hbar = 10^{-27}\,(erg*s) \\ \omega_0 \sim 10^{15}\,(s^{-1}) \\ c = 3*10^{10}\,(cm/s) \\ \tau_2 = 100\,fs = 10^{-10}\,(s) \\ \mu_{QD} = 2.5*10^{-17}\,(cgse) \end{cases} \qquad (4.13)$$

rough estimation gives $n \sim 6,5*10^2$, which is impossible because of -1<n<1. Nevertheless, for order of magnitude higher frequencies the numerical values of $n$ becomes three orders of magnitude lower and requirement n<1 could be satisfied. If it is so, then expression (4.12) assumes that even for a single molecule under some strong enough pump one can get a coherent radiation i.e. set of regularly spaced in time photons. The physical picture behind this expression is following: a generated photon through the self-action produces a stimulated emission causing laser-like auto-oscillation operation mode. This picture is evidently wrong. A spontaneous emission cannot cause at the same time self-stimulated emission; in other words, a spontaneously emitted photon cannot cause a stimulated emission of the next photon from the same quantum system. Moreover, the spontaneous emission is already included into consideration by two relaxation times in (4.7). Note, that for the case of linear HO model this problem does not appear – the emitted photon does not produce nonlinear "self-action" but rather produce linear reaction, which results in extra (radiative) losses in final expressions. The demonstrated paradox can be resolved e.g. by removing the part of the Green function, which is responsible for the spontaneous generation, namely $G_{NF} = \frac{c^2}{4\pi\omega^2}\frac{2}{3}ik^3$. The reason for this modification is in the fact, that the spontaneous emission is already included into the model by two relaxation times and should not be included twice. Hence, as a conclusion: **in the frame of DM formalism the free space part of the Green function has to be artificially excluded from the full expression for the Green function.**

It is worth noting, that this in fact fully corresponds to the case of classic oscillator. In both cases, the stochastic interaction is packed into the relaxation time: in the case of classic oscillator there is only one time $\tau = \frac{1}{\gamma}$, and in case of quantum – two relaxation times $\tau_1$ and $\tau_2$.

Radiative losses in the case of classic oscillator can be equivalently described by coupling with Helmholtz equation or by another term in the dynamic equation. In the case of quantum dipole the radiative losses is already included in the relaxation times and consequently the respective field has to be excluded from the right sides of both classics dynamic equation (3.9) and quantum one (3.10). $A_{NF}$ in (4.7) is whatever but NOT the spontaneous photons, which is mathematically manifested by the absence of term $\frac{c^2}{4\pi\omega^2}\frac{2}{3}ik^3$ in the near field Green function. Math prove behind this statement is given in Appendix 2.

### 4.2 Relaxation dynamics in free space

The relaxation dynamics is described by the same system of equations (3.9) for the classic HO but without any external energy sources and assuming that the frequency ω is now complex value:

$$\begin{cases} p = \beta_c(\omega)\chi A_{NF} \\ A_{NF} = \frac{4\pi\omega^2}{c^2} G_{NF} p \end{cases} \quad (4.14)$$

Substituting the second equation in (4.14) into the first one, we get:

$$p\left(1 - \frac{4\pi\omega^2 \chi}{c^2}\beta_c(\omega)G_{NF}(\omega)\right) = 0$$

$$\Downarrow \quad (4.15)$$

$$\frac{4\pi\omega^2 \chi}{c^2}\beta_c(\omega)G_{NF}(\omega) = 1$$

In other words, we become an equation for the eigen values $\omega_{eigen}$. The imaginary part of the eigen frequency describes the relaxation, namely:

$$\tau_{relax} = \frac{-1}{\text{Im}\left[\omega_{eigen}\right]} \quad (4.16)$$

The relaxation time, described by (4.16) includes all possible influences in the frame of the elaborated model and is also universal i.e. remains valid in the case of NP in the near field zone. The imaginary part of eigen frequency $Im[\omega_{eigen}]$ is always negative and the relaxation time is always positive. The relaxation in free space is obtained by substituting $G_{NF}$ from (4.3) and $\beta_c(\omega)$ from (4.1) into (4.15):

$$\frac{2}{3}i\left(\frac{\omega}{c}\right)^3 \chi \frac{1}{\omega_0^2 - \omega^2 - 2i\gamma\omega} = 1 \Rightarrow$$
$$\Rightarrow \frac{2i\chi}{3c^3}\omega^3 + \omega^2 + 2i\gamma\omega - \omega_0^2 = 0 \tag{4.17}$$

Without radiative losses $\chi = 0$ it is reduced to the known quadratic equation for a HO:

$$\omega^2 + 2i\gamma\omega - \omega_0^2 = 0 \tag{4.18}$$

with simple solution:

$$\omega = -i\gamma \pm \sqrt{\omega_0^2 - \gamma^2} \tag{4.19}$$

According to (4.16) the relaxation time is:

$$\tau_{relax} = \frac{-1}{Im[\omega_{eigen}]} = \frac{1}{\gamma} \tag{4.20}$$

The relaxation processes in quantum emitters require more careful consideration. According to the general rules, relaxation for the quantum oscillator is described by the first equation in (3.10) with $n = -1$ and $\frac{i\mu^2 \tau_2}{\hbar}\frac{4\pi\omega^2}{c^2}G_{NF}^* = 0$, namely:

$$i\tau_2(\omega-\omega_{21})+1=0 \Rightarrow \omega_{eigen}=\omega_{21}+\frac{i}{\tau_2} \qquad (4.21)$$

and the relaxation time is (in quantum case the sign is opposite in compare with (4.17)):

$$\tau_{relax}=\frac{1}{\text{Im}[\omega_{eigen}]}=\tau_2 \qquad (4.22)$$

Note, that this is so called phase relaxation time i.e. averaged time between phase jumps in the quantum emitter oscillation. It has to be realized, that this time is NOT the average time between spontaneous photon emission; the latter is given by $\tau_1$. The relaxation dynamics is described by (3.2) without fields in the right side:

$$\begin{cases} \dfrac{dp}{dt}+p\left(\dfrac{1}{\tau_2}+i(\omega-\omega_{21})\right)=0 \\ \dfrac{dn}{dt}+\dfrac{(n-n_0)}{\tau_1}=0 \end{cases} \qquad (4.23)$$

Radiative losses are already included in the relaxation times, see (3.3), (3.4). The spontaneous emission rate is:

$$P_{FF,Q}(\omega_{12})=\hbar\omega_{12}\frac{n_0}{\tau_{1,r}} \qquad (4.24)$$

Note, that bandwidth of spontaneous emission is not determined in the same mean as for a single photon.

Relaxation rates are usually calculated using both approaches, namely stationary states and by the consideration of the relaxation dynamics. Let us compare these two approaches and consider first stationary case for classic (4.4) and for quantum (4.24) dipoles. Consider a stationary state of the pumped emitter, which is placed near the NP in order to tune the emission rate through the Purcell effect. The emitted power in (4.4) and (4.24) can be presented as photon energy multiplied an effective rate. In both cases (4.4) and (4.24) the measured value is *P*, which depends on the pump $|F|^2$ and $n_0$ in (4.4) and (4.24) respectively. The pump in turn depends on the experimental conditions, which means, that the extractable parameter is not just emission rate, but rather emission rate multiplied by pump. In many papers the pump in the respective expressions are assumed to be constant and independent on e.g. distance between the emitter and NP, which is evidently wrong. One can conclude that in this kind of experiments the emission rate cannot be in general independently extracted.

Let us now compare two expressions for the emitted power (4.4) and (4.24). Let us also assume that according to the Appendix 1 we use a HO equation (A1.4) or (A1.5), i.e. we substitute $\gamma = \frac{1}{\tau_2}$ in (4.4):

$$P_{FF,C}(\omega_0) = \frac{2\pi^2 \omega_0^2 \tau_2^2}{c^3} \frac{|F|^2}{\left(1 + \frac{\chi \omega_0^2 \tau_2}{6c^3}\right)^2} \qquad (4.25)$$

which is evidently not similar to (4.24). It proves again, that these two approaches (classic and quantum) do not give the same results. The reason is in the fact, that classic case (4.4) and respectively (4.25) represents a mix of stochastic relaxation process described by $\gamma$ and regular radiative losses, while quantum case (4.24) is caused only by stochastic interactions.

Finally, let us compare relaxation dynamic equations (4.17) and (4.23) with the "gedanken experiment", namely we assume that the pump is turned off abruptly, and we measure the emitted power in case of classic, or collect statistics in quantum cases. The variation of the relaxation time is provided as before by Purcell effect i.e. by an NP near the emitter. First, the solution of (4.17) is evidently not the same as (4.4), which proves again, that these two tests have to be considered as different ones. Second, (4.17) is free from pump values, and hence can be used to measure the respective relaxation: fitting of the relaxation curves for real and imaginary parts of ω gives both $\gamma$ and $\chi$.

As for the quantum dynamics, the relaxation curve will be proportional to a full (not just radiative or nonradiative) relaxation time, namely:

$$P_{FF,Q}(t) = \hbar\omega_{12} \frac{n_0(t=0)}{\tau_{1,r}} \exp\left(-\frac{t}{\tau_1}\right) \qquad (4.26)$$

and the extractable parameter is $\tau_1$ ($\frac{1}{\tau_1} = \frac{1}{\tau_{1,r}} + \frac{1}{\tau_{1,nr}}$), but not $\tau_{1,r}$. The latter could be also extracted by the fitting of amplitude of (4.26), namely $\hbar\omega_{12} \frac{n_0(t=0)}{\tau_{1,r}}$, but in this case initial pump values $n_0(t=0)$ will be also different and hardly estimated.

The extracted information thus depends crucially on the model, namely classic or quantum one. In the case of a classic model relaxation parameter $\gamma$, pump value $|F|^2$, and coupling parameter $\chi$ can be extracted. This takes place due to the fact that we basically have three independent measurements for three mentioned above values, namely measurement of the central frequency shift (real part of $\omega$), decay rate (imaginary part of $\omega$), and amplitude of the emitted power $P_{FF,C}$.

In contrast, in quantum case there are only two measured values, namely $\frac{n_0}{\tau_{1,r}}$ and $\frac{1}{\tau_1} = \frac{1}{\tau_{1,r}} + \frac{1}{\tau_{1,nr}}$, and three unknowns, namely $n_0$, $\tau_{1,r}$, and $\tau_{1,nr}$. It means that in quantum case it is principally impossible to extract all of these three values. In publications, authors often assume that $n_0$ (pump) remain the same and extract numerical values based on this evidently wrong assumption.

## 5. Conclusions

It is shown that the widely accepted HO equation cannot be used to describe dynamics of a quantum emitter. From the other side, a formal substitution of a HO equation (classic dipole) by DM equations leads to unphysical results. This paradox appears in the case of description of e.g. relaxation dynamics (including Purcell effect) and is expected to appear in spaser dynamics in the case of multimode operation. This paradox can be resolved by careful subdivision of the considered fields by regular and stochastic ones, which is actually a prerequisite for the DM equations. It is shown, that the results of relaxation rates obtained from the consideration of the stationary states differ from the results of relaxation dynamics. It is shown, that the independent extraction of the radiative and non-radiative relaxation times in the quantum model (density matrix) is hardly possible in commonly accepted tests.

## 6. Appendixes

### 6.1 Appendix 1: DM versus HO equation

Dynamics of 2-level system is described by the following system of equations:

$$\begin{cases} \dfrac{d\rho_{12}}{dt} - i\omega_{21}\rho_{12} + \dfrac{\rho_{12}}{\tau_1} = -\dfrac{iH_{12}(\rho_{22}-\rho_{11})}{\hbar} \\ \dfrac{d\rho_{22}}{dt} + \dfrac{\rho_{22}}{\tau_2} = -\dfrac{iH_{12}(\rho_{12}-\rho_{12}^*)}{\hbar} + W \\ \rho_{11} + \rho_{22} = 1 \end{cases} \quad (A1.1)$$

Here ρ describes population (diagonal) and polarisation (non diagonal) dynamics, W is a pump rate, and H is a hamiltonian of interaction (for example, for the interaction with an external electric field it becomes $H_{12} = -\mu * E$.

Sometimes it is more convenient to reduce the dynamics to other variables (Bloch equations)

$$\begin{cases} \dfrac{dQ}{dt} + \dfrac{Q}{\tau_1} - i\omega_{21}P = -\dfrac{2iH_{12}G}{\hbar} \\ \dfrac{dP}{dt} + \dfrac{P}{\tau_1} = i\omega_{21}Q \\ \dfrac{dG}{dt} + \dfrac{1+G}{\tau_2} = -\dfrac{2iH_{12}Q}{\hbar} + 2W \end{cases} \quad (A1.2)$$

Here $P = \rho_{12} + \rho_{12}^*$, $Q = \rho_{12} - \rho_{12}^*$, $G = \rho_{22} - \rho_{11}$

Taking Q from the first equation and substituting it into the second one, the system in Bloch variables takes the following form

$$\begin{cases} \dfrac{d^2 P}{dt^2} + \dfrac{2}{\tau_1}\dfrac{dP}{dt} + \left(\dfrac{1}{\tau_1^2} + \omega_{21}^2\right)P = \dfrac{2\omega_{21}H_{12}G}{\hbar} \\ \dfrac{dG}{dt} + \dfrac{1+G}{\tau_2} = -\dfrac{2H_{12}}{\hbar\omega_{21}}\left(\dfrac{dP}{dt} + \dfrac{P}{\tau_1}\right) + 2W \end{cases} \quad (A1.3)$$

It is clear that the first equation in (A3) is rather far from a trivial HO one. Nevertheless, if it is assumed that G is not changing too much, then the equations become identical:

$$\dfrac{d^2 P}{dt^2} + \dfrac{2}{\tau_1}\dfrac{dP}{dt} + \left(\dfrac{1}{\tau_1^2} + \omega_{21}^2\right)P = \dfrac{2\omega_{21}H_{12}G_0}{\hbar} \quad (A1.4)$$

For example, in case of interaction with the electric field and low intensity (low intensity means that G can be substituted by just -1 corresponding to the situation with all molecules on the lower level $\rho_{11}$)

$$\dfrac{d^2 P}{dt^2} + \dfrac{2}{\tau_1}\dfrac{dP}{dt} + \left(\dfrac{1}{\tau_1^2} + \omega_{21}^2\right)P = \dfrac{2\omega_{21}\mu}{\hbar}E \quad (A1.5)$$

The second possibility to reduce the system to a HO equation is to assume, that due to the pump $W$ the population difference $G$ is kept constant. In order to prove it mathematically, we have to realize that the first term in right side of the second equation in (A3) is a multiplication of two fast oscillating functions, while $G$ has much slower dynamics. The math procedure is in substitution both $P$ and $H$ (or, in case of an electric field $E$) by an anzatz with fast oscillating parts, namely:

$$\begin{aligned} E(t) &= \dfrac{1}{2}\left(A(t)^* \exp(i\omega t) + A(t)^* \exp(-i\omega t)\right) \\ P(t) &= \dfrac{1}{2}\left(p(t)^* \exp(i\omega t) + p(t)^* \exp(-i\omega t)\right) \end{aligned} \quad (A1.6)$$

and then reduce the system (A3) into one equation for slowly varying component p. In this case we have to accept basically Slowly Varying Approximation for the slow amplitudes p and A, which gives us equation for p, but with the first derivative. In this case the system (A3) anyway cannot be reduced to a HO equation.

But let us just assume that due to some reasons the G is kept on some constant level. In this case equation (A4) becomes:

$$\frac{d^2 P}{dt^2} + \frac{2}{\tau_1}\frac{d P}{dt} + \left(\frac{1}{\tau_1^2} + \omega_{21}^2\right)P = -\frac{2\omega_{21}\mu G_0}{\hbar}E \qquad (A1.7)$$

and positive G (inversion) should give us an amplification effect, in this case

$$\frac{d^2 P}{dt^2} + \frac{2}{\tau_1}\frac{d P}{dt} + \left(\frac{1}{\tau_1^2} + \omega_{21}^2\right)P = -\frac{2\omega_{21}\mu |G_0|}{\hbar}E \qquad (A1.8)$$

This equation by no means can be equivalent to the equation of HO with negative absorption. Absorption remains the same (first derivative with some coefficient giving typical relaxation time), and effect of amplification is in a term in right side of the equation (A8), which is equivalent to a kind of external force.

Conclusion:

1. An absorption coefficient in a HO equation can only have a sign, corresponding to an energy losses
2. Rigorous quantum description can be reduced to a HO equation only in case of low, unsaturated losses in a passive (unpumped) quantum system
3. Quantum dynamics, associated with the amplification process (A3) can NOT be described by a HO equation with an inverted sign of losses anyway
4. Equation (A8), which is actually a HO equation with an external force, can be taken to some extend as an equation for polarizability, but even this equation can NOT be obtained through a rigorous math from the DM approach

There are basically no reasons to use a HO equation for amplification, the system (A1) is not too complicated and can be analyzed analytically.

## 6.2 Appendix 2: Maxwell equations and density matrix formalism

A starting point for the consideration is the Helmhoz equation:

$$\Delta A + k^2 A = \frac{4\pi\omega^2}{c^2} P$$

$$P = \mu\left(\langle\psi_2|\mu A|\psi_1\rangle + \langle\psi_2|\mu A|\psi_1\rangle^*\right)$$

$$H\psi = E\psi \tag{A2.1}$$

$$H = H_0 + \mu A;\ H_0\psi_k = E_k\psi_k$$

$$A = A_{regular} + A_{zero\ fluctuatins} + A_{stochastic} \Rightarrow$$

$$\Rightarrow H = H_0 + \mu\left(A_{regular} + A_{zero\ fluctuatins} + A_{stochastic}\right) + V_{stochastic} = H_0 + \mu A_{regular} + \mu\left(A_{zero\ fluctuatins} + A_{stochastic}\right)$$

$A_{zero\ fluctuations}$ are the zero (vacuum) fluctuations of the electric field.

The last scopes in the last equation for $A$ are the combination of the interaction with the vacuum fluctuations and thermo bath. Both are considered as stochastic functions with zero averaged values. As for the Helmholz equation:

$$\Delta A + k^2 A = \frac{4\pi\omega^2}{c^2} P$$

$$A = A_{regular} + A_{zero\ fluctuatins} + A_{stochastic} \Rightarrow \left(\Delta + k^2\right)A_{regular} + \left(\Delta + k^2\right)\left(A_{zero\ fluctuatins} + A_{stochastic}\right) = \frac{4\pi\omega^2}{c^2} P \Rightarrow \tag{A2.2}$$

$$\left(\Delta + k^2\right)A_{regular} = \frac{4\pi\omega^2}{c^2} P\left(A_{regular}\right)$$

$$\left(\Delta + k^2\right)\left(A_{zero\ fluctuatins} + A_{stochastic}\right) = 0$$

Substituting A into the equation for polarization we see, that the stochastic part gets canceled:

$$P = \mu\left(\langle\psi_2|\mu A|\psi_1\rangle + \langle\psi_2|\mu A|\psi_1\rangle^*\right)$$

$$A = A_{regular} + A_{zero\ fluctuatins} + A_{stochastic} \Rightarrow P = \mu\left(\langle\psi_2|A_{regular}|\psi_1\rangle + \langle\psi_2|A_{regular}|\psi_1\rangle^*\right) +$$
$$+\mu\left(\langle\psi_2|(A_{zero\ fluctuatins} + A_{stochastic})|\psi_1\rangle + \langle\psi_2|(A_{zero\ fluctuatins} + A_{stochastic})|\psi_1\rangle^*\right) = \quad (A2.3)$$
$$= \mu\left(\langle\psi_2|A_{regular}|\psi_1\rangle + \langle\psi_2|A_{regular}|\psi_1\rangle^*\right)$$

$$(\Delta + k^2)A_{regular} = \frac{4\pi\omega^2}{c^2}P(A_{regular})$$

As for the Schrödinger equation, it gets transferred into the DM equations, where both $A_{zero\ fluctuatins}$ and $A_{stochastic}$ (i.e. interaction with vacuum fluctuations and with thermo bath) are packed into the two relaxation times $\tau_1$ and $\tau_2$.

$$H\psi = E\psi$$
$$H_0\psi_k = E_k\psi_k$$
$$H = H_0 + \mu\left(A_{regular} + A_{zero\ fluctuatins} + A_{stochastic}\right) = H_0 + \mu A_{regular} + \mu\left(A_{zero\ fluctuatins} + A_{stochastic}\right)$$

$$\begin{cases} \dfrac{dp}{dt} + p\left(\dfrac{1}{\tau_2} + i(\omega - \omega_{21})\right) = \dfrac{i\mu^2 A_{regular}^* n}{\hbar} \\ \dfrac{dn}{dt} + \dfrac{(n - n_0)}{\tau_1} = \dfrac{i(A_{regular}p - A_{regular}^* p^*)}{2\hbar} \\ n = \rho_{22} - \rho_{11} \\ p = \mu\rho_{12} \\ n_0 = \dfrac{(W\tilde{\tau}_1 - 1)}{(W\tilde{\tau}_1 + 1)} \end{cases} \quad (A2.4)$$

In DM equations the both relaxation times depend on the environment according to the local density of state which in turn depends on imaginary part of the respective Green function.